\documentclass{article}
\usepackage{amsfonts}
\usepackage{amssymb}
\usepackage{amsmath}
\usepackage{amsthm}
\usepackage[english]{babel}
\begin{document}

\title{\bf On the role of constraints\\ and degrees of freedom \\ in the Hamiltonian formalism}

\author{{\bf Alexey Golovnev}\\
{\small {\it Centre for Theoretical Physics, The British University in Egypt,}}\\
{\small \it El Sherouk City, Cairo 11837, Egypt}\\
{\small agolovnev@yandex.ru}}
\date{}

\maketitle

\begin{abstract}

Unfortunately, the Hamiltonian mechanics of degenerate Lagrangian systems is usually presented as a mere recipe of Dirac, with no explanation as to how it works. Then it comes to discussing conjectures of whether all primary constraints correspond to gauge symmetries, and it goes all the way to absolutely wrong claims such as the statement that electrodynamics or gravity have only two physical components each, with others being spurious. One has to be very careful because non-dynamical, or constrained, does not mean unphysical. I give a pedagogical introduction to the degenerate Hamiltonian systems, showing both very simple mechanical examples and general arguments about how it works. For the familiar field theory models, I explain why the gauge freedom there "hits twice" in the sense of producing twice as many first-class constraints as gauge symmetries, and why primary, and only primary, constraints should be put into the total Hamiltonian.

\end{abstract}

\section{Introduction}

Hamiltonian mechanics is a very nice formulation for the whole plethora of dynamical models, both for classical and quantum physics, as well as for mathematics since it allows to naturally work with a system of first-order differential equations. In case of non-degenerate Lagrangians, we get unconstrained Hamiltonian systems which express the time-derivatives of all their canonical variables as  functions of the variables themselves. Therefore, we need one and only one Cauchy datum for every canonical variable in order to solve the equations, at least locally.

In every good textbook on classical mechanics, it is explained how the Hamiltonian formalism of non-degenerate systems works. However, when we face degenerate systems, we have to be more careful and to also introduce some constraints on the values of momenta. Those are called primary constraints, and they reduce the number of needed Cauchy data because some combinations of canonical variables get to strictly vanish. 

And here comes another advantage of the Hamiltonian analysis. Since all equations are of the first derivative order, in order to find the number of Cauchy data, or the degrees of dynamical freedom, we simply need to know how many combinations of canonical variables are subject to such restrictions. There are some limitations to this statement though. First, there might be a gauge freedom which makes some variables simply not determined by the equations at all. We call them unphysical and care only about gauge-invariant quantities. Second, the restrictions might be more numerous than the primary ones. The equation of a primary constraint staying zero all the time could demand that some other combinations of variables also vanish, and we call those requirements secondary constraints. Third, not always do Hamiltonian constraints have a clear meaning in the Lagrangian setting. For example, if a variable was totally absent from the Lagrangian, it would have an arbitrary dependence on time, however its momentum would be constrained to be zero. I will discuss all these issues.

Degenerate systems are very important in modern theoretical physics. Indeed, when we have a gauge symmetry, it certainly means that the Lagrangian does not depend on some combination(s) of the fundamental variables in any way, neither through coordinates nor through velocities. Therefore, the system is degenerate and we have primary constraint(s). Very commonly, and I will explain why, such gauge-related constraint produces another constraint. For example, electrodynamics has one gauge freedom and two constraints, while gravity has four gauge freedoms and eight constraints.

Unfortunately, the procedure of working with such systems is usually explained as just a recipe which, as a matter of fact, works well. It works indeed. However, then comes a strange conjecture from the lectures of Dirac \cite{Dirac} that every "first-class", to be defined below, constraint corresponds to a gauge freedom. It is taken very seriously in most of the literature, even though, if taken literally, it is simply wrong. In a sense, the familiar secondary first-class constraints are indeed the consequences of the gauge symmetry, however they do not increase the number of these symmetries, nor possible gauge choices. The number of gauge freedoms is the number of primary first-class constraints only, not more, and even this amount is not always realised as a usual gauge symmetry, due to some other types of singular behaviour of Lagrangians, as I also explain below.

Following the wrong conjecture, as well as putting also the secondary constraints into the Hamiltonian, leads people to claiming that there are only two physical variables in Maxwell electrodynamics, and the same in General Relativity. It is certainly true that those indeed have this number of dynamical modes, but it does not justify calling other non-gauge variables spurious. Doing so they basically say that Coulomb forces and Newtonian potentials are unphysical which is a bit too much to say. In this paper, I first show how the Hamiltonian analysis works in the simplest mechanical cases, though also mentioning some modern physics issues, then I explain why it works and how it must be used, and finally I review its application to electromagnetism and gravity in a bit more detail.

\subsection{A brief review of Hamiltonian mechanics}

I will denote the dependence of a Lagrangian on coordinates and velocities as
$$L(\mathfrak{x,y,z})=L(x,\dot x,y,\dot y,z,\dot z)$$
which will mean that $L$ is taken as a Lagrangian for a system of all the mentioned variables, even if it does not depend on some or all of them. For every Lagrangian variable, the momentum is defined as
\begin{equation}
\label{momentum}
\pi_x\equiv\frac{\partial L}{\partial\dot x}.
\end{equation}
In cases when the matrix $\frac{\partial^2 L}{\partial\dot x_i \partial \dot x_j}$ is non-degenerate, the Lagrangian system of $L({\mathfrak x}_i)$ is called non-degenerate, too, and this definition (\ref{momentum}) of $\pi=\pi(x, \dot x)$ can be inverted to $\dot x=\dot x (\pi,x)$, at least locally.

Then we introduce the Hamiltonian function
\begin{equation}
\label{HamFun}
{\mathcal H}(x_i,{\dot x}_i,\pi_{x_i})=\sum_i \pi_{x_i} {\dot x}_i - L({\mathfrak x}_i).
\end{equation}
For a non-degenerate system, and up to a possible non-uniqueness of solving for velocities in terms of momenta, we can finally define the canonical Hamiltonian as
\begin{equation}
\label{CanHam}
\left. \begin{array}{cc}
H(x_i,\pi_{x_i})={\mathcal H}(x_i,{\dot x}_i(\pi_{x_i},x_i),\pi_{x_i})=\left(\mathop{\sum}\limits_i \pi_{x_i} {\dot x}_i - L({\mathfrak x}_i)
\right) \\ \vphantom{.}
\end{array} \right|
\begin{array}{cc}
\vphantom{ABCD} \\ {\dot x}_i : \quad \pi_{x_i}=\frac{\partial L}{\partial{\dot x}_i}
\end{array}.
\end{equation}

One can easily find the differential equations satisfied by the Hamiltonian by simply looking at the variation of the Hamiltonian function (\ref{HamFun}) restricted to the surface defined by imposing the equation for momenta (\ref{momentum}):
\begin{equation}
\label{variation}
\left. \begin{array}{cc}
\delta H =\delta \left(\mathop{\sum}\limits_i \pi_{x_i} {\dot x}_i - L({\mathfrak x}_i)
\right) \\ \vphantom{.}
\end{array} \right|
\begin{array}{cc}
\vphantom{ABCD} \\ \pi_{x_i}=\frac{\partial L}{\partial{\dot x}_i}
\end{array}
\begin{array}{cc}
= \mathop{\sum}\limits_i\left( {\dot x}_i \delta\pi_{x_i} -\frac{\partial L}{\partial x_i}\delta x_i\right)\\
\vphantom{.}
\end{array}
\end{equation}
which implies
$$\frac{\partial H}{\partial \pi_{x_i}}=\dot x_i \qquad \mathrm{and} \qquad \frac{\partial H}{\partial x_i}=-\frac{\partial L}{\partial x_i}.$$

Of course, the first equation simply solves for velocities in terms of momenta, while the second one can then be compared with the Lagrangian equation by observing that the latter tells us that
$$\frac{\partial L}{\partial x_i}=\frac{d}{dt}\frac{\partial L}{\partial {\dot x}_i}={\dot\pi}_{x_i},$$
and therefore the Hamiltonian shape of the equations of motion is
\begin{equation}
\label{eqHam}
\dot x_i=\frac{\partial H}{\partial \pi_{x_i}} \qquad \mathrm{and} \qquad {\dot\pi}_{x_i}=-\frac{\partial H}{\partial x_i}.
\end{equation}
For a convex Lagrangian, it can be simply viewed as the new Legendre transform, which is involutive, and therefore brings us back to the Lagrangian with the variational principle written in terms of $\int\left(\pi \dot x - H\right)$.

The equations of motion (\ref{eqHam}) can be presented in the beautiful language of symplectic geometry, 
$${\dot f}(x,\pi)=\{f,H\},$$ 
such that the canonical coordinates and momenta are the Darboux coordinates for the (anti-symmetric) Poisson bracket: $\{x_i,\pi_{x_j}\}=\delta_{ij}$, with brackets among coordinates or among momenta being zero. Therefore, we have a natural symplectic geometry on the phase space defined by the Poisson bracket of two functions $f(x,\pi)$ and $g(x,\pi)$:
\begin{equation}
\label{Poisson}
\{f,g\}\equiv \sum_i \left(\frac{\partial f}{\partial x_i} \cdot \frac{\partial g}{\partial \pi_{x_i}} - \frac{\partial f}{\partial \pi_{x_i}} \cdot \frac{\partial g}{\partial x_i}\right).
\end{equation}
Here I will not go into any more detail on that. There are many nice and rigorous expositions of the Hamiltonian mechanics for non-degenerate systems, see for example the book by Arnold \cite{Arnold}.

\subsubsection{Non-quadratic kinetic Lagrangians -- Hamiltonians with branches}

Let me just briefly mention that it all works with no surprise if the Lagrangian is a positive-definite quadratic form of velocities. Having higher powers of velocities brings its own issues. Of course, even if the Hessian  $\frac{\partial^2 L}{\partial\dot x_i \partial\dot  x_j}$ is non-degenerate, it might have some singular phase space points with vanishing determinant. On top of that, even in the regular parts, the solution $\dot x = {\dot x}(\pi, x)$ is often non-unique.

Just as a very simple example, consider $L(\mathfrak x)=\frac13 {\dot x}^3$. The momentum is $\pi={\dot x}^2$, and it is an even function of $\dot x$. The Hamiltonian is easily found to be $H=\pm \frac23 \pi^{3/2}$, and its two branches correspond to positive and negative velocities. This example is very trivial, however there are other options. For instance, one can also play with time crystals \cite{TCr}.

\subsubsection{Degenerate systems -- the Dirac's recipe}

Now I turn to the main topic of this paper: Hamiltonian systems with constraints. We call a Lagrangian degenerate if its Hessian $\frac{\partial^2 L}{\partial {\dot x}_i \partial {\dot x}_j}$ has vanishing determinant, and therefore it has a non-trivial kernel and cannot be inverted. Such systems cannot be written as ${\ddot x}_i=f_i(x,\dot x) $.

In this case, the definition (\ref{momentum}) of the momenta in terms of velocities is not invertible either. The variations of velocities are mapped into variations of momenta by action of the Hessian matrix. Assuming some smoothness and constancy of rank, the dimension of its kernel gives the number of independent velocity variations which do not change the momenta at all. On the other hand, this dimension is the codimension of its image in the momentum space, and the equation of the image submanifold is the set of the primary constraints, in the number equal to the dimension of that kernel. Let's define the primary constraints $\Phi_a$ as equations 
$$\Phi_a(\pi_{x_i},x_i)=0$$ 
to be put on top of the usual Hamiltonian equations, and the index $a$ running from $1$ to the dimension of the Hessian's kernel. In other words, the primary constraints are necessary and sufficient (again, at least locally) conditions for a set of momenta $\pi_{x_i}$ to be in the image of the mapping (\ref{momentum}).

We can define the Hamiltonian function (\ref{HamFun}) the same way again. The sad news is that we cannot find the velocities as functions of momenta in order to substitute them there. However, the calculation of its variation (\ref{variation}) goes the same way as before, the only difference is that the assumption of the eq. $(\ref{momentum})$ put into this variation generically cannot work, for any velocity, unless we always put the momentum onto the primary constraint manifold. However, it's not what we can control when taking derivatives of the Hamiltonian once we have forgotten the initial origin of this function.

Moreover, we cannot even find a unique canonical Hamiltonian with the usual property (\ref{variation}) of the first variation. Indeed, if we change it as $H\longrightarrow H +  \mathop{\sum}\limits_a \Phi_a \cdot {\mathcal F}_a (x_i, \pi_{x_i})$ with arbitrary functions ${\mathcal F}_a$, it does not change the equation (\ref{variation}) because in the case of imposing the condition (\ref{momentum}), both on the background and on the variation, we always have $\Phi_a=0$ . Therefore, the next natural step is to take an arbitrary possible choice of the canonical Hamiltonian and define the total Hamiltonian
\begin{equation}
\label{TotHam}
H_T = H + \sum_a \lambda_a \Phi_a
\end{equation}
with arbitrary functions $\lambda_a$ called Lagrange multipliers.

The main recipe is that the equations are then
\begin{equation}
\label{equations}
\dot x_i=\frac{\partial H_T}{\partial \pi_{x_i}} \qquad \mathrm{and} \qquad {\dot\pi}_{x_i}=-\frac{\partial H_T}{\partial x_i}
\qquad \mathrm{and\ the\ constraints} \qquad \Phi_a=0.
\end{equation}
Dirac preferred to call the last equality a weak one and to denote it as $\Phi_a\approx 0$. It is weak in the meaning that it is valid on the constraint surface only, and not when one goes away from it, and in particular the first two equations do get non-trivial terms from differentiating the $\Phi$-s. For me, it is just yet another equation which I do not want to distinguish anyhow. On the contrary, if I want to stress that something is identically zero, I use the sign of $\equiv 0$. After all, it is all but absolutely common that even when a function takes the  zero value, its derivative might be non-vanishing. Interestingly enough, Dirac in his paper \cite{inDirac} denoted weak equalities with the usual equality sign, while for strong equalities he used the identical equality notation, even though his definition of a strong equality there was rather that of ${\mathcal O}(\epsilon^2)$ as opposed to ${\mathcal O}(\epsilon)$ type.

My claim about the system of equations (\ref{equations}) is that, in every reasonable case, this is already the final answer, and nothing else is needed in order to have the equations identical to the Lagrangian ones. One way to look at this is that we are doing the variation (\ref{variation}) always respecting the condition that the momenta are only those which indeed fulfill the condition (\ref{momentum}) and therefore do correspond to at least some velocity. I will later give more detailed explanations. However, intuitively one can see it as varying $\int (p\dot x - H + \lambda \Phi)$, with the last term both imposing the necessary requirements on the momenta and reproducing every possible canonical Hamiltonian.

Let me stress it once more. The equations (\ref{equations}) with primary, and only primary, constraints are all what we need. But when we want to easily and reliably count the degrees of freedom, we need to be more careful.

All in all, when I take $\Phi_a=0$ as an equation of motion, it immediately implies that $\dot\Phi_a=0$, too. Taken together with other equations, it might or might not entail new constraints, $\tilde\Phi_m$, which are called secondary. Using the other equations for that  can also be rephrased in terms of Poisson brackets (\ref{Poisson}):
$$0=\dot\Phi_a=\{\Phi_a, H_T\}$$
which takes the form of a function of coordinates and momenta, with no time-derivatives of those. Then we can do the same with the secondary constraints, to possibly get tertiary ones and so on... I will call all of them secondary $\tilde\Phi_m$. At every step, if the Poisson bracket of a constraint with the total Hamiltonian is not either identically zero or expressed in terms of the constraints themselves (weakly zero in the language of Dirac), it means either a restriction on the Lagrange multipliers or yet another constraint.

Once all the constraints are found, they reduce the number of independent Cauchy data, while those Lagrange multipliers of primary constraints which haven't got fixed in this process count unphysical variables in the system. Here comes classification of constraints. We can compute the whole matrix of Poisson brackets of all constraints with each other. If every element of this matrix is "weakly zero", i.e. the constraints form a closed algebra with respect to the brackets, we say that they are first-class. If the matrix is, to the contrary, non-degenerate on the constraint surface, then they are second-class. If the matrix is neither of that, one can separate the constraints, maybe after some redefinitions, into first- and second-class ones.

If all the constraints are first-class, then the conditions of $\{\Phi_a, H_T\}=0$ and $\{\tilde\Phi_m, H_T\}=0$ can not be satisfied by conditions on the Lagrange multipliers. Therefore, in this case every primary constraint goes with an arbitrary $\lambda$ in the Hamiltonian equations of motion (\ref{equations}), normally representing a gauge freedom. If a primary constraint led to a secondary one with which it is of second-class, then it seems that its Lagrange multiplier gets fixed by preservation of its secondary friend. Since every constraint kills a Cauchy datum, while first-class ones are also associated with gauge symmetries, the common lore says that each first-class constraint and each pair of second-class constraints reduce the number of degrees of freedom by one.

\subsubsection{My comments}

In this paper I argue that it is not that simple, and it is better to always check what happens in each particular case, even though the Hamiltonian formalism can sometimes be more transparent than the Lagrangian one, indeed. My main point is that the amount of gauge freedom is normally reflected in the number of only primary first-class constraints. And even this statement requires certain regularity of the Lagrangian. It can happen that some variables enter the Lagrangian only being multiplied by a factor which vanishes  together with its first variation on every configuration in which other variables satisfy all the equations. Then the former variables will enjoy the full unphysical freedom of being arbitrary functions of time, at the same time without any off-shell symmetry of such Lagrangian. I will give an example below.

In what concerns the Dirac conjecture, the relation between first-class constraints and gauge symmetries is not trivial. In a sense, all of them are normally related to some gauge symmetry. However, in non-trivial cases when the "gauge hits twice", not even the primary ones are the precise generators of the gauge transformations, with this role being played by some combination of them with the secondary one(s) \cite{Castellani}. If we take the conjecture as a statement that all first-class constraints are independent generators of gauge transformations, then it is simply wrong, in all the cases which have secondary first-class constraints.

There is one more point about that. It is claimed sometimes \cite{Teitelboim} that it would be better to add all the constraints to the Hamiltonian, not only primary but also secondary ones, in order to get what is called an extended Hamiltonian,
\begin{equation}
\label{ExtHam}
H_E = H_T + \sum_m \tilde\lambda_m \tilde\Phi_m = H + \sum_a \lambda_a \Phi_a + \sum_m \tilde\lambda_m \tilde\Phi_m.
\end{equation}
First, it is not needed. The primary constraints are already enough to reproduce the Lagrangian equations of motion in the Hamiltonian language. Second, at least with the first-class constraints, it is simply wrong. In this case, adding a secondary constraint to the Hamiltonian also adds a new arbitrary Lagrange multiplier producing a fake effective gauge symmetry on top of the real one.

\section{Very simple examples of systems with constraints}

I will first illustrate the workings of constrained Hamiltonian formalism with very simple examples, then give some general mechanical statements, and finally discuss the usual field theories we have in physics.  There are several good expositions of Hamiltonian mechanics with constraints \cite{Teitelboim,Gitman,Rothe}, however more as a recipe, and the important subtleties are rarely paid enough attention to.

\subsection{Trivial gauge symmetry}

Let me start from a very simple example
$$L(\mathfrak x)=0$$
which has one variable $x(t)$ and a gauge symmetry which kills it. Finally, we have nothing physical, one gauge freedom and zero degrees of freedom. With the same success, I could have taken $L=1$, for anyway its variation under arbitrary $\delta x$ is zero.

\subsubsection{Failure of forgetting to put primary constraints into the Hamiltonian}

We get the zero momentum $\pi_x=0$ which is its only (first-class) constraint $\Phi\equiv \pi_x$, and the  Hamiltonian function (\ref{HamFun}), ${\mathcal H}=\pi_x {\dot x} - L$, is equal to zero on the constraint surface for any choice of $\dot x$. If we simply take
$$H=0,$$
then the equations of motion (\ref{eqHam}) tell us that
$${\dot x}=0 \qquad \mathrm{and} \qquad \dot{\pi}_x=0.$$
It gives then higher freedom of the momentum than in the initial Lagrangian formulation, but at the same time requires more from the variable $x$ which used to be absolutely free. Even adding the equation of $\pi_x=0$ by hand would not help to restore the arbitrariness of $x(t)$. A primary constraint is not a mere selection of a surface inside a phase space, it genuinely changes the dynamics. Therefore, we must use the primary constraints in the Hamiltonian.

And it is also a first sign of the need for being careful when calculating the number of degrees of freedom in the Hamiltonian formulation. If I assume that a zero Hamiltonian $H\equiv 0$ for $n$ variables has no constraints, it means $2n$ required Cauchy data, and works like $n$ degrees of freedom. However, half of these data simply set the momenta to arbitrary constant values which have no influence on the coordinates. For finding all the coordinates we need only $n$ Cauchy data which set their otherwise arbitrary constant values, and then it goes like $n$ halves degrees of freedom. In particular, as we saw above, in a theory with one variable it gives $\frac12$ degree of freedom.

\subsubsection{Zero Hamiltonian with a primary constraint}

Now, let us correctly take the primary constraint into account and construct the total Hamiltonian (\ref{TotHam})
$$H_T=H+\lambda\Phi=0+\lambda \pi_x$$
with an arbitrary Lagrange multiplier $\lambda$. The new equations (\ref{equations}) are
$${\dot x}=\lambda \qquad \mathrm{and} \qquad \dot{\pi}_x=0 
\qquad \mathrm{with\ the\ constraint} \qquad \pi_x=0.$$
And now we are totally back to the Lagrangian situation. The momentum is fully fixed by the primary constraint, and the full freedom in the behaviour of the coordinate got restored by the Lagrange multiplier being an arbitrary function. The equation ${\dot x}=\lambda$ still requires one Cauchy datum to solve it for $x$ in terms of $\lambda$, but it makes no real sense since the Lagrange multiplier is assumed to be an absolutely arbitrary function of time, no extra condition restricts it; and the same is finally true of $x$.

Note that we actually could have taken an absolutely different canonical Hamiltonian there. Of course, $H=0$ is indeed equal to the function ${\mathcal H}=\pi_x \dot x - L$ at the surface of $\pi_x=0$ with $L=0$. However, any canonical $H$ of the form $H=\pi_x\cdot f(x,\pi_x)$ would do the same. And the total Hamiltonian $H_T= \pi_x\cdot f(x,\pi_x) +\lambda \pi_x$ produces then equivalent equations of motion. In other words, the canonical Hamiltonian (\ref{CanHam}) for a degenerate system is not unique. And actually, if it was not for the primary constraint put into $H_T$, these choices would have given different equations of motion.

If for the model being trivial, the choice of $H=0$ looks as the only natural one above, we could also take a somewhat less trivial case of $L(\mathfrak{x,y})=\frac12 \left({\dot y}^2 - y^2\right)$ with the Hamiltonian $H_T=\frac12 \left(\pi_y^2 + y^2 \right) + \lambda \pi_x$ yielding the usual harmonic oscillator of $y(t)$ and total freedom for $x(t)$ again the same way as before. It is still very natural, to not add any contribution of $\pi_x$ to the canonical Hamiltonian. A bit funnier way, though, is to take 
$$L(\mathfrak{x,y})=\frac12 \left(\left({\dot x}-{\dot y}\right)^2 - \left(x-y\right)^2\right).$$
In this case, the function ${\mathcal H}=\frac12 \left(\left({\dot x}-{\dot y}\right)^2 + \left(x-y\right)^2\right)$ can be represented as $ H=\frac12\left( \pi^2_x + (x-y)^2\right)$, or $H=\frac12\left( \pi^2_y +(x-y)^2\right)$, with only symmetry considerations suggesting either $H=\frac14\left( \pi^2_x + \pi^2_y\right) +\frac12 (x-y)^2$ or $H=-\frac12 \pi_x \pi_y +\frac12 (x-y)^2$. Looking at the $\frac{x\pm y}{\sqrt{2}}$ variables offers yet another symmetric choice of $H=\frac18\left( \pi_x - \pi_y\right)^2 +\frac12 (x-y)^2$. Those are all the same at the surface of the primary constraint $\pi_x+\pi_y=0$, with the same equations of motion produced by $H_T=H + \lambda\left( \pi_x+\pi_y\right)$.

\subsubsection{Adding total time derivatives}

Note that Lagrangians different by only a total time derivative are equivalent, at least at the level of equations of motion. Therefore, a gauge transformation might also be preserving a Lagrangian up to a total time derivative, or a Lagrangian density up to a total divergence, in field theories. 

Therefore, another option to take the model with empty physical content is 
$$L({\mathfrak x})=f(x)\dot x.$$
We get then a non-zero momentum $\pi_x = f(x)$, the natural choice of zero canonical Hamiltonian, and $H_T=\lambda (\pi_x - f(x))$. The equations, together with the constraint, then read
$$\dot x =\lambda, \qquad \dot\pi_x = f^{\prime}\cdot\lambda, \qquad \pi_x=f.$$
They are obviously satisfied by an arbitrary $x(t)$ and $\pi_x(t)=f(x(t))$.

For the Lagrangian variable $x(t)$, the result is of course the same as it was with the zero Lagrangian. However, the Hamiltonian description goes in a bit different way, due to the different definition of the momentum, and therefore the constraint $\Phi=\pi -f$ instead of $\Phi=\pi$. It shows that if a gauge symmetry is realised only up to a total derivative term, the Hamiltonian representation of it might well go in a more elaborate style. One good field theory example of it is the Lorentz symmetry \cite{Lor} of the Teleparallel Equivalent of General Relativity (TEGR).

\subsection{A very simple gauge symmetry can also hit twice}

There exists a simple slogan \cite{Teitelboim, MV} that "a gauge symmetry hits twice" meaning that it produces two Hamiltonian constraints. As we have seen in the trivial example, or as can be found in the Refs. \cite{Teitelboim, MV} too, it is not always true. And I will come to it later, but let me also mention now that another example is TEGR. If both its diffeomorphism and Lorentz symmetries hit twice, then using the usual naive counting, it would have $16-2\cdot 4 - 2\cdot 6=-4$ degrees of freedom in four spacetime dimensions.

However, one can easily construct a simple mechanical model in which a gauge hits twice indeed:
$$L(\mathfrak{x,y})=\frac12 \left({\dot x} + y\right)^2.$$
We get the momenta $\pi_x=\dot x +y$ and $\pi_y=0$, the latter being the primary constraint $\Phi=\pi_y$. Then the total Hamiltonian (\ref{TotHam}) can be written as
$$H_T=\frac12 \pi_x^2- y \pi_x +\lambda \pi_y,$$
with all the possible additions of $\pi_y$ to the canonical part if one wants to write the same story in a slightly more complicated form. The equations of motion (\ref{equations}) are
$$\left\{\begin{array}{cc}
 {\dot x}=\pi_x - y \vphantom{\int_{\int}} \\
 \vphantom{\int^{\int}}   {\dot y} = \lambda
\end{array}\right.
 \qquad \mathrm{and} \qquad \left\{\begin{array}{cc}
 {\dot \pi}_x=0 \vphantom{\int_{\int}} \\
 \vphantom{\int^{\int}}    {\dot \pi}_y = \pi_x 
\end{array}\right. 
\qquad \mathrm{with\ the\ constraint} \qquad \pi_y=0.$$
These equations, or the formal "preservation" of the primary constraint, yield the secondary constraint $\tilde\Phi\equiv \pi_x =0$ which in turn is trivially preserved due to one of the equations, or to the fact of $H_T$ not depending on $x$. Then what we get is $\dot x = -y$ with an arbitrary $y(t)$. It seems we have one pure gauge variable, two first-class constraints, and one Cauchy datum needed, therefore $\frac12$ degrees of freedom. Note however that the same system of equations can be interpreted as an arbitrary $x$ and $y=-\dot x$ which then means no Cauchy data at all. The very naive counting looked like $\frac12$, but in reality there is $0$ dynamical degrees of freedom, even though not both variables are pure gauge. The physical combination of variables is $\dot x + y$, and it is fully constrained to be equal to zero. The traditional counting of dynamical modes by subtracting the number of first-class constraints works well in this case.

The Lagrangian story is of course the same. The equations of motion are
$$\frac{d}{dt}(\dot x + y)=0 \qquad \mathrm{and } \qquad \dot x + y=0.$$
The first one is a dynamical equation from variation of $x$, and the second one is a constraint, the only constraint of the Lagrangian formalism. The action has a gauge symmetry under 
$$x\longrightarrow x+ u(t),\qquad  y\longrightarrow y - \dot u(t)$$
so that only one combination of the two variables, $\dot x + y$, is physical. However, the primary constraint $\Phi=\pi_y$ transforms only $y$ leaving $x$ intact which is not a gauge transformation. And this is the reason of the secondary constraint appearing. In other words, it is the same as what happens in electrodynamics: the gauge symmetry mixes a velocity (time derivative) with another coordinate (other spatial derivatives) producing two first-class constraints by only one gauge symmetry.

Note that the variation of $y(t)$ in the action has produced the only Lagrangian constraint $ \dot x + y=0$, with another combination of the two variables being pure gauge. This constraint corresponds to the secondary Hamiltonian constraint, while the primary one has no relation to the Lagrangian variables, it is simply $\pi_y=0$ which in itself does not restrict any velocity. In the Lagrangian language, we have one gauge freedom which strips one combination of variables of physical meaning, while the physical one is subject to a constraint. If we choose a gauge of $x=0$, the physical equation (not a gauge choice!) tells us that $y=0$ which means zero Cauchy data. At the same time, the gauge of $y=0$ has a remaining symmetry of constant $u$ which is reflected in the solution of an arbitrary constant $x$, not contributing to the physical gauge-invariant variable $\dot x + y$. In the Hamiltonian language, we have both momenta constrained, while for the two coordinates there is one arbitrary Lagrange multiplier, so that naively the system needs one Cauchy datum. The origin of this number is in the fact that the Hamiltonian equations look in such a way as if taking the gauge in terms of $y$ was more natural.

\subsubsection{Beware of secondary constraints put into the Hamiltonian}

It is often stated, for example in the classic book of Henneaux and Teitelboim \cite{Teitelboim}, that it is more natural to take both primary and secondary constraints on equal footing. This is totally wrong! First, it is not natural even from the basic framework of the Hamiltonian formalism. The primary constraints are the restrictions on possible values of momenta. They might lead to some other restrictions on the variables, called secondary constraints, but those aren't new restrictions, they directly follow from the Hamiltonian equations and the primary constraints (\ref{equations}). Second, adding the secondary constraints to the total Hamiltonian (\ref{TotHam}) gives an extended Hamiltonian (\ref{ExtHam}) which often changes the physics of the system.

In particular, for our model we get
$$H_E=H_T+\tilde\lambda \tilde\Phi=\frac12 \pi_x^2- y \pi_x +\lambda \pi_y + \tilde\lambda \pi_x,$$
with the new equations of motion
$$\left\{\begin{array}{cc} 
{\dot x}=\pi_x - y +\tilde\lambda \vphantom{\int_{\int}} \\
 \vphantom{\int^{\int}}   {\dot y} = \lambda\end{array}\right. 
\qquad \mathrm{and} \qquad \left\{\begin{array}{cc}
 {\dot \pi}_x=0 \vphantom{\int_{\int}} \\ 
\vphantom{\int^{\int}}    {\dot \pi}_y = \pi_x \end{array}\right. 
\qquad \mathrm{with\ the\ constraints} \qquad \left\{\begin{array}{cc}  
\pi_y=0 \vphantom{\int_{\int}} \\
 \vphantom{\int^{\int}} \pi_x=0 
\end{array}\right. .$$
As also in the correct approach, we have both momenta equal zero; however now the two independent Lagrange multipliers make both Lagrangian variables arbitrary functions of time. In other words, we have spoiled the correspondence with the Lagrange equations. The total Hamiltonian was equivalent to the initial Lagrangian system with one gauge freedom and one physical variable, though a constrained one. At the same time, the extended Hamiltonian introduced one more arbitrary Lagrange multiplier which resulted in a new model with two unphysical freedoms and not any single physical variable, thus rather corresponding to $L(\mathfrak{x,y})=0$.

Note also that primary and secondary constraints are not the same, even if the number of Lagrange multipliers is kept correct. If we had put only $\tilde\Phi$ in our Hamiltonian, it would have produced constant $\pi_y$ but not necessarily zero, and constant $y$ with an arbitrary $x(t)$, a totally different system.

\subsection{Trivial second-class constraints}

The simplest example of a second-class constraint pair would be
$$L(\mathfrak x)=-x^2.$$
It immediately has a constraint of $\pi_x=0$ and the naturally possible canonical Hamiltonian $H=x^2$.

The total Hamiltonian is $H_T=x^2 + \lambda \pi_x$. Its equations $\dot x=\lambda$ and $\dot\pi_x=x$ together with the primary constraint $\pi_x=0$ reproduce the Lagrangian equation $x=0$ which is also the secondary constraint. In this case, the Lagrangian system has one single constraint $x=0$ which, being of non-derivative character, is enough for reducing the number of Cauchy data to zero. The Hamiltonian representation does the same by having two second-class constraints. At the same time, its Lagrange multiplier is fixed to $\lambda=0$.

Note that incorrectness of adding the secondary constraint to the Hamiltonian in the previous example was due to another arbitrary Lagrange multiplier changing the level of predictability. It is not necessarily the case with the second-class constraints since all the multipliers are constrained then. For example, obviously $H_E=x^2 +\lambda \pi_x +\tilde\lambda x$ reproduces the same Lagrangian equation of $x=0$ and demands $\lambda=\tilde\lambda=0$. 

It happened because both the coordinate and its momentum are strictly fixed by the pair of constraints which can not be changed by adding any new terms to the Hamiltonian. Related to that, and as a simple exercise about the non-uniqueness of the Hamiltonian, one can check that changing the canonical Hamiltonian (\ref{CanHam}) in this case from $x^2$ to, say, $\pi_x^2 + x^2$ would change neither the $H_T$ nor the $H_E$ results, while taking it as $\pi_x + x^2$ would shift the multiplier $\lambda$ by $1$ with no change to physics either. 

I will not discuss second-class constraints any longer. Let me only mention that introducing Dirac brackets is often taken as a must for the Hamiltonian analysis. This is of course not true. Their purpose is different. If we want to canonically quantise a system with second-class constraints by promoting the Poisson brackets to commutators of operators, then we cannot impose all the constraints in terms of operators, not even as the physical subspace of the Hilbert space belonging to the kernels of all the constraints, unless there is no physics and its subspace is empty. Here come Dirac brackets, as a part of an algorithm for quantisation, which we don't need in this paper.

\subsection{First-class without a gauge}

Now I take another Lagrangian
$$L(\mathfrak{x,y})=\frac12 y{\dot x}^2$$
which gives the primary constraint of $\pi_y=0$ and the total Hamiltonian (\ref{TotHam})
$$H_T=\frac{\pi_x^2}{2y}+\lambda \pi_y.$$

The Lagrangian equations of motion were
$$y\ddot x + \dot x \dot y=0 \qquad \mathrm{and} \qquad {\dot x}^2=0$$
implying an arbitrary constant value of $x$ and an absolutely arbitrary $y(t)$. Hamiltonian equations (\ref{equations}) yield
$$\left\{\begin{array}{cc} 
{\dot x}=\frac{\pi_x}{y} \vphantom{\int_{\int}} \\
 \vphantom{\int^{\int}}   {\dot y} = \lambda
\end{array}\right. \qquad \mathrm{and} 
\qquad \left\{\begin{array}{cc}
 {\dot \pi}_x=0 \vphantom{\int_{\int}} \\
 \vphantom{\int^{\int}}    {\dot \pi}_y = \frac{\pi_x^2}{2y^2} 
\end{array}\right. 
\qquad \mathrm{with\ the\ constraint} \qquad \pi_y=0$$
with the same result for the Lagrangian variables while requiring both momenta to be zero.

Note that the secondary constraint of $\pi_x=0$ followed from the primary constraint and the Hamiltonian equation for $\pi_y$. Of course, it is nothing but requiring that the primary constraint is preserved in time. In terms of required Cauchy data, it is one half of a degree of freedom, not zero like the traditional counting would suggest. At the same time, the set of constraints,
$$\Phi=\pi_y \qquad \mathrm{and} \qquad \tilde\Phi=\pi_x,$$
is a system of two first-class constraints, with no gauge freedom in the Lagrangian model.

This model was also considered in the form of $L=e^y {\dot x}^2$ in the book \cite{Teitelboim} by Henneaux and Teitelboim. Their claim is that it has a gauge freedom for $y$ but not for $x$, and therefore is a counterexample to Dirac's conjecture. My point is that the Dirac's conjecture is actually never correct. So, I don't argue against this conclusion. However, in this case there is simply no gauge freedom in the Lagrangian at all. It shows that sometimes a primary first-class constraint might be unrelated to any gauge freedom. 

Why they call it a gauge system in the book \cite{Teitelboim} is because an arbitrary $y(t)$ satisfies the equations of motion. However, in this case, it is not due to a symmetry of the Lagrangian, not even up to a total derivative. It is due to another kind of singularity here: the variable $y$ enters the Lagrangian being multiplied by a square of something which is equal to zero when equations of motion are satisfied, and therefore its contribution to the equations does not identically disappear but rather gets multiplied by something which is zero due to another equation. 

If the Lagrangian was of the form $L(\mathfrak{x,y})=f(y)g^2(x,\dot x)$ with whatever smooth functions $f$, with $f^{\prime}\neq 0$, and $g$, it would be precisely the same story with the equations of motion satisfied by any function $y(t)$ and a function $x(t)$ which is a solution of the equation $g=0$. Indeed, $y$ is then arbitrary. But I won't call it a gauge symmetry. It shows that the unphysical nature of a variable, in the sense of not being determined by equations of motion, might sometimes be coming not from a gauge freedom.

On the other hand, this non-gauge freedom is indeed a kind of singularity I mentioned. If we had taken a Lagrangian $L=y\dot x$, it would have been a model with two constant in time Lagrangian variables governed by a zero canonical Hamiltonian and a pair of second-class constraints. From the Cauchy data counting, it has one degree of freedom, though it rather looks like two halves. An analogous result would be with $L=y\left({\dot x}^2 - x^2\right)$ when $x \neq 0$, or in a Lorentz-invariant field model like the mimetic fluid \cite{mim1,mim2} $\lambda\left((\partial_{\mu}\phi)(\partial^{\mu}\phi)-1\right)$, for in these cases the extra variable is multiplied in the Lagrangian by something which does have a non-vanishing first variation around the physical trajectory. Needless to say, if one wants to have it singular again, we can always put an extra square like in the $fg^2$ model mentioned above.

\subsubsection{On bifurcations, new Mimetic, teleparallel craze, and all that}

I guess that the main reason why Henneaux and Teitelboim \cite{Teitelboim} took the previous example with $e^y$ instead of just $y$ is in order to avoid a point with zero $\pi_x$. Actually, in the case of $L=y{\dot x}^2$ the variable $y$ being zero does not change anything in the equations of motion. The Hamiltonian looks singular when $y\to 0$, however everything works fine if we simply ignore the fact that $\frac{1}{y}$ might become infinite. On the other hand, one could try to consider an extra primary constraint of $\pi_x=0$ when $y=0$. If applied naively, it gives a zero canonical Hamiltonian with both momenta being primary constraints, and therefore producing an incorrect result of arbitrary $x(t)$.

Actually, the reason of nothing special at $y=0$ is that the momentum $\pi_x$ as a function of $y$ had only a simple root which was not enough for it acting as a fundamental constraint. If we had $L(\mathfrak{x,y})=y^2 {\dot x}^2$, then the locus of zero $y$ would be a special place with a bifurcation: the function $x(t)$ is indeed absolutely arbitrary there. All the $y\neq 0$ regimes can then be described as above with substitution of $y$ by $y^2$, while the story at $y=0$ goes like a zero Hamiltonian with constraints for both momenta and $y$ being set to zero.

There is no general understanding of how to work with these bifurcations at the Hamiltonian level. Though usually it happens only at a very specific point which we are trying to avoid as pathological, for example like a locus of $f^{\prime}=0$ in $f(R)$ gravity. More problematic is that, even though the bifurcations appearing at the primary level are usually quite obvious right away, they can also come at the later stages, and in a less immediately visible manner. Namely, the nature of constraints can be changing.

As yet another example of a non-gauge singular situation at the Lagrangian level, let's consider a Lagrangian 
$$L(\mathfrak{x,y})=\frac12 \left( {\dot x}^2 - x^2 (y-x)^2\right).$$ 
Its total Hamiltonian with the primary constraint $\Phi=\pi_y$
$$H_T= \frac12  \left( {\pi_x}^2 + x^2 (y-x)^2\right) +\lambda \pi_y$$ 
produces the secondary constraint $\tilde\Phi=x^2 (y-x)$. Their Poisson bracket is $\{\tilde\Phi, \Phi\}=x^2$. Therefore, at $x\neq 0$ we have a pair of second-class constraints, while at $x=0$ it turns into two first-class constraints. The very obvious Lagrangian equations show that indeed these regimes are genuinely different. We either have $x=0$ and totally arbitrary $y(t)$, or $y=x$ and the dynamical equation of $\ddot x=0$.

A more familiar case of that is the Proca theory. When we have a vector field with the Lagrangian density $ L(\mathfrak A)=-\frac14 F_{\mu\nu} F^{\mu\nu} + V(A_{\mu}A^{\mu})$ with $F_{\mu\nu}\equiv \partial_{\mu} A_{\nu} - \partial_{\nu} A_{\mu}$, the usual $U(1)$ gauge symmetry is broken by the potential term $V$, and therefore all four components are physical with three of them being dynamical and $A_0$ being constrained by a pair of second-class constraints. However, if there exists a point with $V^{\prime}=0$, then the gauge invariance is spontaneously restored for linear perturbations around it, and the constraints get to have their Poisson bracket vanishing. It effectively brings us back to the gauge-invariant case with only three physical components and two of them dynamical. Therefore, the number of degrees of freedom appears to not be well-defined.

Maybe, the most cautious attitude would be to simply avoid such situations. However, sometimes people want to use them on purpose. For example, mimetic gravity \cite{mim1, mim2} was built by a metric transformation $\tilde g_{\mu\nu} = g^{\alpha\beta}(\partial_{\alpha}\phi)(\partial_{\beta}\phi) g_{\mu\nu}$ from the General Relativity (GR). If the physical metric is $\tilde g_{\mu\nu}$ while $g_{\mu\nu}$ and $\phi$ are used as fundamental variables, then the role of conformal mode is played by the scalar field which, due to more derivatives, gives more solutions in the shape of an effective ideal fluid \cite{mim2}. Very many different modifications have been considered \cite{mimrev, Slava2}. However, one can also stick to the simple initial idea and generalise it to disformal transformations which then produce the same mimetic gravity as long as the metric transformation is not invertible, otherwise it is pure GR with a spurious scalar field on top \cite{Nathalie}. Recently, it was noticed \cite{mimnew1, mimnew2} that in most cases the metric transformation is invertible, except at some singular loci which then produce extra solutions of mimetic type, too. Personally, I am skeptical about using such ill-posed theories \cite{mimnewme}, but it would be interesting to learn how to work with them.

Another amazing story of recent years is modified teleparallel gravity such as $f(T)$. It breaks the local Lorentz invariance in the space of tetrads (I assume its pure tetrad formulation), with however a rather chaotic zoo of remnant symmetries \cite{remnant}. The structure of the constraint algebra is so difficult that there is even no consensus on the number of degrees of freedom \cite{telprobl}. What is almost fully clear is that there is some new dynamics on top of the usual two modes of GR \cite{telprobl, nontrmin}. However, the trivial Minkowski space is an obvious strong coupling regime with the Lorentz symmetry restoration, and therefore no new modes in the linear perturbations. What is much more surprising is that no new dynamical modes are found around the usual cosmologies either \cite{meTomi, cosmMan}, even though the Lorentz symmetry is broken there and the new features of cosmological perturbation theory arise, such as non-zero gravitational slip \cite{meTomi}. It might be the case that such theories do require a genuinely novel insight into the Hamiltonian methods.

\subsubsection{A brief remark on Proca-Nuevo de nuevo}

Let me also mention that, inspired by ghost-free massive gravity \cite{mgr,revClaudia}, yet another vector field theory has been proposed \cite{Claudia}, with the Lagrangian built in terms of elementary symmetric polynomials of the eigenvalues of a square root of the matrix $\delta^{\mu}_{\nu} + \partial^{\mu} A_{\nu} + \partial_{\nu} A^{\mu} + \partial^{\mu} A_{\alpha}  \partial_{\nu} A^{\alpha}$. It has a primary constraint which sounds promising since dynamical temporal  components are usually deadly. Very recently, a new investigation of the model appeared \cite{newProca}, though only in 2D. Lagrangian analysis of constraints \cite{LagrCount} was used to claim that the primary constraint is the only one in the story. Then  assuming that there must be the same number of Hamiltonian constraints, and that it cannot be first-class in case of no gauge symmetry, the conclusion was that it is a case of a constraint which is second-class with itself, and therefore the model has $\frac32$ degrees of freedom \cite{newProca}. 

A self-second-class constraint is a rare and exciting situation, known for example in versions of Ho{\v r}ava gravity. In my opinion, it is a very interesting topic; however, given the subtleties of the dictionary between the Lagrangian and the Hamiltonian formalisms I reviewed above, it would be very important to do an explicit Hamiltonian analysis, as well as to get more Lagrangian details. Of course, the structure of the model is not very simple, but in the case of massive gravity the full Hamiltonian analysis was successfully done  by brute force approach \cite{mgr1, secmgr}, and some other methods of working with square roots of matrices are also known \cite{mgr2, mgr3}. 

\section{Hamiltonian mechanics with constraints}

Now let me discuss what happens in general, mainly for Lagrgangians quadratic in velocities. My main goal here is to show that  in general a total Hamiltonian is enough to reproduce the Lagrangian physics. For demonstrating the fact that insisting on extended Hamiltonians is incorrect, it is enough to have some explicit examples, in the previous and in the next Sections.

Let's take a Lagrangian
\begin{equation}
\label{exsys}
L(\mathfrak{x,y})=\frac12 {\dot x}^2 + b(x,y) \dot x +c(x,y) \dot y + d(x,y)
\end{equation}
which has the momenta $\pi_x=\dot x + b$ and $\pi_y=c$. It produces the primary constraint $\Phi=\pi_y - c(x,y)$, and the total Hamiltonian (\ref{TotHam}) can be chosen as
$$H_T=\frac12 (\pi_y - b)^2 -d+ \lambda \left(\pi_y - c\right).$$
And it is only for clarity of presentation that I take just two variables. It won't change much of what follows, if I take $x$ as an $n$-dimensional vector, and $y$ -- as an $m$-dimensional one.

The Lagrangian equations of motion are easily found to be
$$\left\{\begin{array}{cc} 
\ddot x =\left(\frac{\partial c}{\partial x} - \frac{\partial b}{\partial y} \right)\cdot \dot y + \frac{\partial d}{\partial x} \vphantom{\int_{\int}} \\  
\vphantom{\int^{\int}}  0=\left(\frac{\partial b}{\partial y} - \frac{\partial c}{\partial x} \right)\cdot \dot x + \frac{\partial d}{\partial y} 
 \end{array}\right.$$
while the Hamiltonian ones (\ref{equations}) are
$$\left\{\begin{array}{cc} 
\dot x = \pi_x - b \vphantom{\int_{\int}} \\  
\vphantom{\int^{\int}} \dot y = \lambda
 \end{array}\right. \qquad \mathrm{and} 
\qquad \left\{\begin{array}{cc}
 {\dot \pi}_x= (\pi_x - b)\frac{\partial b}{\partial x} + \frac{\partial d}{\partial x} + \lambda \frac{\partial c}{\partial x} \vphantom{\int_{\int}} \\
 \vphantom{\int^{\int}}    {\dot \pi}_y= (\pi_x - b)\frac{\partial b}{\partial y} + \frac{\partial d}{\partial y} + \lambda \frac{\partial c}{\partial y}
\end{array}\right. 
\qquad \mathrm{with\ the\ constraint} \qquad \pi_y - c=0.$$
Differentiating $\dot x = \pi_x - b$, we get $\ddot x = \dot\pi_x - \frac{\partial b}{\partial x} \dot x - \frac{\partial b}{\partial y} \dot y$ which upon substitution of the equation for $\dot\pi_x$, and taking into account $\pi_x-b=\dot x$ and $\lambda= \dot y$, gives the first Lagrangian equation. Analogously, differentiation of the constraint, $\dot\pi_y= \frac{\partial c}{\partial x} \dot x + \frac{\partial c}{\partial y} \dot y$, transforms the equation for $\dot\pi_y$ into the second Lagrangian equation. Therefore, the Hamiltonian method works well indeed.

What has happened here is the following. As we discussed in the Introduction, for non-degenerate systems the canonical Hamiltonian (\ref{CanHam}) has a very nice property (\ref{variation}) of $\frac{\partial H}{\partial \pi}=\dot x$ and $\frac{\partial H}{\partial x}= -\frac{\partial L}{\partial x}$, then the former equation means simply that $\pi=\frac{\partial L}{\partial \dot x}$ while $\dot\pi=-\frac{\partial H}{\partial x}$  comes back to the Lagrangian equation of $\frac{d}{dt}\frac{\partial L}{\partial \dot x}=\frac{\partial L}{\partial x}$. Degenerate systems are more intricate. As long as we keep the momenta strictly equal to their Lagrangian definition, the variation of the canonical Hamiltonian is still given by $\delta H=\dot x \delta\pi - \frac{\partial L}{\partial \dot x}\delta x$. However, once we have found some corresponding expression $H=H(x,\pi)$ and forgotten about $\dot x$, it fails to be unique and can contain arbitrary combinations of primary constraints.

In the current example, it corresponds to adding arbitrary combinations of $\pi_y - c$ to the Hamiltonian, which we explicitly took into account by adding the primary constraint term in $H_T$. Therefore, the equation for $\dot x$ has correctly found the velocity of $x$ in terms of its momentum, while $\dot y$ remains arbitrary, at least at this stage. The momentum $\pi_y$ did not depend on $\dot y$ which therefore cannot be found from the values of momenta, and the primary constraint term takes care of that by putting a Lagrange multiplier in place of the non-existing $\dot y (\pi_y)$. On the other hand, these additions also make the variations $\frac{\partial H}{\partial x }$ and $\frac{\partial H} {\partial y}$ different from $-\frac{\partial L}{\partial x}$ and $-\frac{\partial L}{ \partial y}$, by the terms with derivatives of $c(x,y)$ which are taken care of by the primary constraint in $H_T$, too.

Therefore, different values of $\lambda$ in $H_T$ correspond to all possible definitions of the canonical Hamiltonian with the correct variation at the primary constraint surface. When we impose also the equation of the primary constraint, $\pi_y - c = 0$ in our case, it comes back to the desired variations. Either a normal to constraint surface variation determines the only correct $\lambda$ which is then like the contact force acting on a particle from the surface on which it moves, and the variable $y$ appears to be physical but constrained, or we have it just arbitrary, and this is then a case of gauge freedom and its induced primary first-class constraint.

Note that we have checked it without ever talking about the secondary constraints. The equations just tell us everything about the Lagrangian variables, and looking at other consequences is needed only for counting the number of degrees of freedom. In particular, if $\frac{\partial b}{\partial y} \equiv \frac{\partial c}{\partial x}$, so that $b\dot x + c \dot y$ is a total time derivative, and $ \frac{\partial d}{\partial y}\equiv 0$, then $y(t)$ is a purely gauge mode, the first Lagrangian equation is the only one and is dynamical; at the same time the Hamiltonian formalism has an arbitrary Lagrange multiplier and one first-class constraint, just for the formal definition of the momentum of the unphysical variable. 

Otherwise, both variables are physical but we are getting the secondary constraint expressed by the second Lagrangian equation. In this case there is a pair of second-class Hamiltonian constraints. The Lagrangian formalism immediately has a constraint for $\dot x$, and substituting it into the equation for $\ddot x$ we get another constraint for $\dot y$. In the Hamiltonian language, the restriction of $\dot y$ comes as a fixed value of the Lagrange multiplier $\lambda$ required by preservation of the secondary constraint. All in all, we have two Cauchy data and therefore one, or two halves, degree of freedom. A special case would be if  $\frac{\partial b}{\partial y} \equiv \frac{\partial c}{\partial x}$ is still satisfied but with $\frac{\partial d}{\partial y}$ not identically zero. Then $y$ is fully constrained to an extremum of $d$ while $x(t)$ is having a fully dynamical equation of motion. 

Note that if $x$ and $y$, as well as $b$ and $c$, were multidimensional, it could potentially cause a complicated constraint algebra, in terms of Poisson brackets, with many possible secondary constraints and so on. However, the formal demonstration that the Lagrangian and the Hamiltonian ($H_T$) equations do tell us the same physics would go with no big change. 

\subsection{A hint of more general considerations}

We have just discussed the Hamiltonian formalism for Lagrangian systems quadratic in velocities. Let me now give some abstract description of how it should work in general. Imagine we have a Lagrangian $L(\mathfrak{x}_i)$ with its equations
$$\frac{\partial^2 L}{\partial {\dot x}_i \partial {\dot x}_j} {\ddot x}_j + \frac{\partial^2 L}{\partial {\dot x}_i \partial x_j} {\dot x}_j - \frac{\partial L}{\partial x_i}=0$$
and are trying to construct a Hamiltonian (\ref{CanHam}) for it.

Let's assume that, around a point of the configuration space we are looking at, there exists a matrix $Q$ such that
$$\frac{\partial^2 L}{\partial {\dot x}_i \partial {\dot x}_j}=Q\cdot
\left(\begin{array}{cc}
\ell & 0\\
0 & 0
\end{array}\right)\cdot Q^{-1}$$
with ${\ell}\neq 0$ and $Q^T=Q^{-1}$, with ${^T}$ now meaning "transpose". Only for getting it less cumbersome have I taken it two-dimensional, and it can be thought of as two multidimensional blocks with $\ell$ being then a non-degenerate matrix.

From the definition of momenta (\ref{momentum}), $\pi_{x_i}\equiv\frac{\partial L}{\partial\dot x_i}$,
we get the following variations of them in terms of the Lagrangian velocities:
\begin{equation}
\label{varvar}
\overrightarrow{\delta \pi} = Q \left(\begin{array}{cc}
\ell & 0\\
0 & 0
\end{array}\right) Q^{-1} \overrightarrow{\delta \dot{x}} + \left(  \frac{\partial^2 L}{\partial {\dot x} \partial x} \right)  \overrightarrow{\delta x}.
\end{equation}
If I denote the projector to the second component as $P_2\equiv  \left(\begin{array}{cc}
0 & 0\\
0 & 1
\end{array}\right)$, then the primary constraint in terms of the variations acquires the form of
\begin{equation}
\label{convar}
P_2 Q^{-1} \left(\overrightarrow{\delta \pi} -  \left(  \frac{\partial^2 L}{\partial {\dot x} \partial x} \right)  \overrightarrow{\delta x} \right)=0
\end{equation}
showing that $QP_2Q^{-1}$ is the projector to the direction in which the momentum $\overrightarrow\pi$ can only be changed due to a change of coordinates and does not feel the velocities. At the same time, in the complimentary direction of $QP_1Q^{-1}$, there is a fixed relation between the momentum and the velocity.

Then, as we have discussed before, even in the degenerate case with the non-invertible relation (\ref{varvar}) between variations, we can anyway build a canonical Hamiltonian (\ref{CanHam}) trying to reproduce the desired variations (\ref{variation}) as
$$\delta H = \mathop{\sum}\limits_i\left( {\dot x}_i \delta\pi_{x_i} -\frac{\partial L}{\partial x_i}\delta x_i\right).$$
What I had swept under the rug is the displeasing question of which precisely velocity stays there. Even if we do only a variation allowed by the constraint, it can actually be any velocity which corresponds to the value of $\overrightarrow\pi$, therefore near our point it is, roughly speaking, something definite in $QP_1Q^{-1}$ subspace, and can be arbitrary in $QP_2Q^{-1}$.

What exactly we get, depends on how we construct the canonical Hamiltonian. Recall that it is not unique. Our usual way was to avoid adding any terms proportional to the primary constraint. For example, in the calculations above with the Lagrangian (\ref{exsys}) we could have added something proportional to $\pi_y - c$ to the canonical Hamiltonian. But we didn't. This was a way to avoid the unwanted velocity in the term $\left(\pi_y - c\right) \dot y$ by just putting it to zero. This way, not only did we expel $\pi_y$ from the canonical Hamiltonian but also got rid of the function $c(x,y)$ in it.

In other words, we construct the canonical Hamiltonian from only the "$QP_1Q^{-1}$-part" of momenta which, together with keeping track of the variational relation (\ref{convar}), means that we are losing some of the $\overrightarrow{\delta x}$ variations. Without trying to give a good mathematical meaning to all that, let me say that we basically get only the $QP_1Q^{-1}$ projection of ${\dot x}_i$ from the $\frac{\partial H}{\partial \pi_{x_i}}$ derivative, while in calculating $\overrightarrow{\frac{\partial L}{\partial x}}$ we first subtract a term like ${\overrightarrow{\dot x}}^T \cdot QP_2Q^{-1}\overrightarrow{\frac{\partial L}{\partial \dot x}}$ from the Lagrangian. In our previous example (\ref{exsys}), we can check it by putting $\lambda=0$ for now in its Hamiltonian equation. Indeed, the $\frac{\partial H}{\partial\pi}$ part gave only the non-trivial value of $\dot x$ having $\dot y=0$, while the $\frac{\partial H}{\partial x}$ equations lack derivatives of $c(x,y)$ since it had dropped off from the canonical Hamiltonian.

What was removed from the equations by this restriction on the canonical Hamiltonian, is restored by adding the $\lambda\Phi$ term to the total Hamiltonian $H_T$. Indeed, the variation of the constraint has the form of the eq. (\ref{convar}), and then it gives an arbitrary $Q P_2 Q^{-1}$ component of the velocity and a $ Q P_2 Q^{-1}\frac{\partial^2 L}{\partial {\dot x} \partial x}$ term to the equation for $\dot \pi$, therefore taking care of the change in variation of $x$ in the canonical piece. Note that this is precisely what happened with $\lambda$-terms in the Hamiltonian equations for the model (\ref{exsys}).

The important point is that, precisely as before, it has restored too much. We've got an arbitrary value of the velocity component corresponding to the kernel of the Hessian. It is expected that the multiplier $\lambda$ either remains arbitrary, which usually shows a case of gauge symmetry, or it will be fixed by equations (and give a secondary second-class constraint). 

Needless to say, what I have given here is not a mathematical proof of the Lagrangian-Hamiltonian correspondence for beyond the quadratic-in-velocities case. For example, we would have to worry about the arguments of the coefficients given by partial derivatives of the Lagrangian, and so on. It is an interesting problem, but I only meant to provide a gist of how it should work.

\section{Our usual field theories}

After having discussed the mechanical cases, which in my opinion are very important for understanding the theoretical foundations and the workings of the method, I turn to the field theories we often use in physics, namely electromagnetism and gravity.

\subsection{Electrodynamics}

I will consider Maxwell electrodynamics in vacuum. Therefore, the action is very simple,
\begin{equation}
\label{EM}
L(\mathfrak A)=-\frac14 F_{\mu\nu} F^{\mu\nu} \qquad \mathrm{with} \qquad  F_{\mu\nu}\equiv \partial_{\mu} A_{\nu} - \partial_{\nu} A_{\mu}\qquad
\mathrm{by\ definition.}
\end{equation}
The momenta (\ref{momentum}) are easily found to be
$$\pi_0=0 \qquad \mathrm{and} \qquad \pi_i=F_{0i}={\dot A}_i-\partial_i A_0,$$
from which we immediately see the primary constraint $\Phi=\pi_0$ and the total Hamiltonian
$$H_T=\frac12 \pi_i^2 + \pi_i \partial_i A_0 + \frac14 F^2_{ij} + \lambda \pi_0.$$
As before, we could have added an arbitrary function of $\pi_0$ to the canonical Hamiltonian, but it is anyway taken care of by the primary constraint term.

Note that the Lagrangian equations of motion are
$$\partial_{\mu} F^{\mu\nu}=0.$$
At the same time, the Lagrangian has a gauge symmetry of $A_{\mu}\longrightarrow A_{\mu}+\partial_{\mu}\psi$, therefore only three components out of four are physical. And still, not all of them are dynamical because one of the equations, the temporal one, is a constraint, $0=\partial_i F_{0i}=\partial_i{\dot A}_i - \bigtriangleup A_0$. Naively, it looks like reducing the number of degrees of freedom by $\frac12$. However, one possible gauge choice is $\partial_i A_i=0$, which can be achieved by the gauge transformation with $\psi$ such that $\bigtriangleup\psi=-\partial_i A_i$, and then the result is a full strength constraint of $\bigtriangleup A_0=0$. We have two dynamical variables then and one more physical but constrained. Of course, the latter corresponds to the Gau{\ss}'s law of $\overrightarrow{\partial}\cdot \overrightarrow E = 0$ in absence of electric charges, with $\overrightarrow E$ being the electric field.

The Hamiltonian equations we get are
\begin{equation}
\label{EMeq}
\left\{\begin{array}{cc}
 {\dot A}_0= \lambda \vphantom{\int_{\int}} \\
 \vphantom{\int^{\int}}   {\dot A}_i = \pi_i + \partial_i A_0
\end{array}\right.
 \qquad \mathrm{and} \qquad \left\{\begin{array}{cc}
 {\dot \pi}_0=\partial_i \pi_i \vphantom{\int_{\int}} \\
 \vphantom{\int^{\int}}    {\dot \pi}_i = \partial_j F_{ji} 
\end{array}\right. 
\qquad \mathrm{with\ the\ constraint} \qquad \pi_0=0.
\end{equation}
Again, it fully reproduces the Lagrangian theory. Indeed, from the primary constraint and the equation for $\dot \pi_0$, which is also simply $\{\Phi,H_T\}$, we get the secondary constraint
$$\partial_i \pi_i =0$$
 which, using the equation for $\dot A_i$, immediately reproduces the Lagrangian constraint. And we easily find the dynamical equations: $$\partial_j F_{ji}={\dot\pi}_i={\ddot A}_i-\partial_i {\dot A}_0={\dot F}_{0i}.$$ 

Incidentally, we see that the full gauge invariance is there: only the combinations of $F_{\mu\nu}$ matter. And again, naively, the equation for $A_0$ looks like requiring one Cauchy datum. But this is of course spurious since it makes no sense to worry about an initial value of something whose derivative is given by an arbitrary function. At a deeper level, this is irrelevant because having fixed the value of $A_0$, we are still allowed to do another gauge transformation with a parameter depending on spatial coordinates only, while if we have fixed the value of ${\dot A}_0$, then we can still have another gauge parameter which is at most linear in time. This symmetry is not so clear from the Hamiltonian representation.  However it is  also there because the vector potential $A_i$ comes only in combinations of $F_{0i}$ and $F_{ij}$, even in the Hamiltonian equations.

\subsubsection{On incorrectness of extended Hamiltonians again}

Let me state it again that extended Hamiltonians are not correct descriptions of gauge theories. In this case we would get
$$H_E=\frac12 \pi_i^2 + \pi_i \partial_i A_0 + \frac14 F^2_{ij} + \lambda \pi_0 + \tilde\lambda \partial_i \pi_i$$
with an extra, and totally unjustified, arbitrary variable $\tilde\lambda$ in the equations. Indeed, the only modification is to one of the equations: 
$${\dot A}_i = \pi_i + \partial_i A_0 + \partial_i \tilde\lambda.$$ 
Now we are allowed to change temporal and longitudinal fields independently, and this is totally wrong. It's not the usual electrodynamics! It is all right for perturbatively quantising photons in vacuum, but it totally neglects Coulomb forces as if those were unphysical.

Indeed, if we accept both Lagrange multipliers, even the electric field is not physical. And we can then produce whatever effective electric charges:
$\partial_i F_{0i}=  \bigtriangleup \tilde\lambda $, out of nothing. Note that the case with $\tilde\lambda=\tilde\lambda(t)$ changes nothing, and this is because of the symmetry of our Hamiltonian: $\tilde\lambda(t) \partial_i \pi_i \to \pi_i \partial_i \tilde\lambda(t)=0$. However, allowing here for two fully arbitrary Lagrange multipliers is wrong. It would be like if I could freely do gauge transformations for temporal and spatial components separately $\left\{
\begin{array}{cc}
A_0 \longrightarrow A_0 + \dot\phi  \\
A_i \longrightarrow A_i + \partial_i \psi
\end{array}
\right.$ with independent $\phi=\int dt \int dt\ \lambda$ and $\psi=\int dt \int dt\ \lambda + \int dt\ \tilde\lambda$, which is not a symmetry of electrodynamics.

In particular, sometimes people say \cite{MV, Rafael} that in electrodynamics we must fix two gauges, $A_0=0$ and $\partial_i A_i=0$. No, this is not a gauge choice! As I showed also above, we cannot arbitrarily change both temporal and longitudinal fields. This choice is possible only in electrodynamics without sources, and it is not only a gauge choice but also a solution of the Lagrangian constraint equation. Indeed, since we have the Gau{\ss}'s law (for vanishing electric charge density)
$$\partial_i{\dot A}_i - \bigtriangleup A_0=0,$$
once we have chosen a gauge of $A_0=0$, we get $\frac{d}{dt}(\partial_i A_i)=0$. Then it can be reduced to $\partial_i A_i=0$ using a gauge parameter $\psi$ depending on spatial coordinates only which is a totally legitimate remnant symmetry then. By no means is it a new gauge choice. It is simply the fact that $A_0=0$ does not fully fix a gauge.

Out of pure curiosity, note also that our total Hamiltonian equations (\ref{EMeq}) have reproduced the initial definition of the spatial momenta: $\pi_i={\dot A}_i - \partial_i A_0$ which is not surprising since the definition of momenta was invertible in its spatial part. This success is totally destroyed by introduction of the extended Hamiltonian. If we demand it by hand though, then the extended Hamiltonian picture comes back to the real physics, and fixes the new Lagrange multiplier to be a function of only time which makes it practically non-existent together with any incorrect effects it has produced. However, there is no natural reason for doing so on the Hamiltonian side.

\subsection{General Relativity}

Another important gauge theory we have is General Relativity. Its particularly convenient representation for the analysis is in ADM variables \cite{ADM}. Those are given in terms of the metric written as
$$g_{\mu\nu}dx^{\mu}dx^{\nu}=(N^2- N_i N_j \gamma^{ij}) dt^2 - 2N_i dt dx^i - \gamma_{ij} dx^i dx^j.$$
Therefore, we use the spatial metric $\gamma_{ij}\equiv -g_{ij}$ and all the spatial indices are operated on by this spatial metric, while all the other metric components are expressed in terms of new variables, the lapse $N$ and  shift $N_i$.

Using the extrinsic curvatures of the constant time slices
$$K_{ij}=\frac{1}{2N} \left({\mathop{\bigtriangledown}\limits^{(3)}}\vphantom{\Gamma}_i N_j+
{\mathop{\bigtriangledown}\limits^{(3)}}\vphantom{\Gamma}_j N_i - \dot{\gamma}_{ij}\right),$$
and after some very straightforward but relatively cumbersome algebra \cite{meADM}, one can find the curvature scalar
$$\sqrt{-g}R=\sqrt{\gamma}N\left({\mathop{R}\limits^{(3)}}+K^{ij}K_{ij}-K^i_i K^j_j\right)
-2\partial_0\left(\sqrt{\gamma}K^i_i\right)+2\sqrt{\gamma}{\mathop{\bigtriangledown}\limits^{(3)}}\vphantom{\Gamma}_j \left(K^i_i N^j\right)
-2\sqrt{\gamma}{\mathop{\bigtriangleup}\limits^{(3)}} N$$
in terms of that for the spatial metric.
Neglecting the total derivatives, we come to the Lagrangian
\begin{equation}
\label{GR}
L(\mathfrak{N, N}_i,\mathfrak{g}_{ij})=\sqrt{\gamma}N\left({\mathop{R}\limits^{(3)}}+K^{ij}K_{ij}-K^i_i K^j_j\right).
\end{equation}

Then we see that $\pi_N=\pi_{N_i}=0$ while $\pi^{ij}\equiv\pi_{\gamma_{ij}}=\sqrt{\gamma}\left(K^k_k \gamma^{ij}-K^{ij}\right)$, and the total Hamiltonian is
$$H_T=-\int d^3 x\sqrt{\gamma}\left(N\left(\mathop{R}\limits^{({\mathit 3})}+\frac{1}{\gamma}\left(\frac12 \left(\pi^j_j\right)^2-\pi_{ik}\pi^{ik}\right)
\right)+2N^i\mathop{{\bigtriangledown}^k}\limits^{({\mathit 3})}\pi_{ik} \right)+\lambda\pi_N+\lambda_i \pi_{N_i}.$$
The secondary constraints are directly obvious from the Hamiltonian. More work is required to check that all the constraints are first-class, which is quite natural though because all the four pairs again come from a gauge symmetry which mixes temporal and spatial derivatives. In other words, the secondary constraints come about because the gauge transformations are not just changes of lapse and shift.

It is very common in the literature to omit the $\lambda$-terms, and to simply treat this Hamiltonian system as the one with zero canonical Hamiltonian, the secondary constraints in the role of primary ones, and the lapse and shift as Lagrange multipliers. In this case, it is actually correct since the equations are not changed then except renaming some field variables into Lagrange multipliers. Indeed, if I change from $H_T=y f(x,\pi_x)+\lambda\pi_y$ to $H_T=\lambda f(x,\pi_x)$, the only difference then is that the equations for $x$ and $\pi_x$ instead of the factor of $y$, which was anyway arbitrary as $\dot y= \lambda$, now get just another arbitrary factor, that of $\lambda$ itself. The only thing to remember then is that the diffeomorphisms got somewhat hidden since they now involve transformations of the new Lagrange multipliers, which is unusual.

Like in the case of electrodynamics, the naive counting of dynamical degrees of freedom works well. From the ten variables, they subtract eight first-class constraints and get two degrees of freedom. What is however wrong is to claim that there are only two physical combinations of the metric components in General Relativity. That would mean to recognise only gravitational waves (GWs) as physical and to neglect all the Newtonian forces. This is not how the world is built. 

The best way to see that there are more physical components is to study the cosmological perturbation theory \cite{MFB}. The coordinate change functions $\delta x^{\mu}$ are all the four independent gauge parameters, and they leave six variables physical. On top of the two graviton polarisations, there are two components of a physical vector and two physical scalars (Bardeen potentials). While the vectors are not very important for cosmology because they are decaying almost inevitably, the scalars play an instrumental role in calculating the CMB observables and the large scale structure. They are not dynamical by themselves, and the acoustic waves appear because of the matter contents of the Universe, but how these waves evolve and lead to structure formation is crucially governed by gravity, and not just in its two GW polarisations.

\subsubsection{Gravity with tetrads -- not every gauge hits twice}

So, as we have seen, the familiar gauge symmetries normally "hit twice" because of mixing different derivatives of different components. At the same time, it is not difficult to construct examples of not hitting twice. In the case of electrodynamics, I could have put $A_{\mu}-B_{\mu}$ everywhere instead of just the simple vector $A_{\mu}$. This is a very trivial symmetry, and it would be associated with four new first-class constraints of $\pi_{A_{\mu}} + \pi_{B_{\mu}}=0$ which would not "hit twice" and would really be the new gauge generators precisely by themselves.

However, in gravity we have a more appealing way to go. Namely, let's write the GR in terms of tetrads. At the same time, I will not introduce any spin connection. What I have in mind is rather teleparallel-type generalisations of general relativity than coupling fermions. We take
$$L({\mathfrak e}^a_{\mu})=-\sqrt{g}R(g) \qquad \mathrm{where} \qquad g_{\mu\nu}\equiv \eta_{ab} e^a_{\mu} e^b_{\nu}$$
and get the gauge symmetry of Lorentz rotations $e^a_{\mu} \longrightarrow \Lambda^a_b e^b_{\mu}$. Since the metric is invariant under such transformations, and with no derivatives of $\Lambda$ involved in them, it is very easy to see that there appear six Lorentz constraints which do not hit twice. They simply subtract the six newly introduced variables (from ten components of the metric to sixteen of the tetrad), and don't do anything else. 

An interesting point is that I was a bit cheating in the paragraph above. It is so simple only if our Lagrangian is indeed written in terms of the metric. However, there is also the Teleparallel Equivalent of General Relativity (TEGR) which is the basis of modified teleparallel theories. The action of TEGR is not given in terms of the metric. Its Lagrangian is locally Lorentz invariant only up to a surface term (which is its difference from usual Lagrangians of GR), and therefore the Lorentz behaviour of the Hamiltonian formalism becomes much more difficult and interesting \cite{Lor}.

\subsubsection{On the canonicity of ADM}

The final topic I want to touch upon is the strange claims that appeared some time ago \cite{strange} and continue to be seen in the current literature \cite{morestrange}. The main statement there is that the ADM formalism is not a canonical Hamiltonian treatment of general relativity, and is not equivalent to Dirac's approach to Hamiltonian relativity. Of course, it simply cannot be true. The only difference between different approaches to Hamiltonian GR is in the boundary terms omitted in order to not have higher derivatives and to not bother with unnecessary complications of Ostrogradsky procedure, and also in the choice of variables.

The ADM variables of course complicate the way diffeomorphisms look, simply because those are different variables. Nevertheless, all the symmetries are there, since the Lagrangian density is still a scalar density, up to a boundary term. Nothing is done there, except rewriting it in less obviously symmetric variables. And Castellani \cite{Castellani}, for example, was able to trace the gauge transformations there. It doesn't matter if we have to parametrise the gauge transformations in a more difficult way for finding them in one procedure or another. One can add any function of canonical variables multiplied by a primary constraint to the canonical Hamiltonian. It won't change any physics, but will reparametrise the Lagrange multiplier. The only real question is whether the symmetry is there. And it is.

However, one point I would like to specifically mention is their claim that ADM variables are not canonical variables for gravity \cite{strange}. First of all, there is nothing sacred about being canonical. If I do a non-canonical transformation, it simply means that my Poisson bracket will be changed, and my Hamiltonian equations get to look less nice. Moreover, one and the same theory can be written with different choices of canonical variables, just having different Hamiltonians. One way to get that is by using boundary terms. For example, Lagrangians $L(\mathfrak{x,y})=\frac12 \left({\dot x}^2+{\dot y}^2\right)$ and $L(\mathfrak{x,y})=\frac12 \left({\dot x}^2+{\dot y}^2\right) + x\dot y + y \dot x$ are equivalent. However, they have different definitions of momenta. Therefore, canonical variables of one of them are not canonical for another one. The difference has no physical implication though, because the Hamiltonians are also different and the final equations for $x$ and $y$ are equivalent.

Therefore, different Hamiltonian formulations of GR might indeed have different symplectic structures, and it does not mean that only one of them is the correct one. On the other hand, they argue so by saying that the change of variables from metric to ADM is not canonical \cite{strange}. And this is not true either. If two theories are obtained from each other by only a change of Lagrangian variables, then their canonical variables are related by a conformal transformation.

It is very simple to explain what went wrong with their calculation of $\{N,\pi^{ij}\}$ in which they got a non-zero result \cite{strange}. If we go from one set of variables to another, the momenta are also changed, even for those variables which are not changed. In particular, ${\dot g}_{00}=2N\dot N - 2N_i {\dot N}_j \gamma^{ij}+N^i N^j {\dot g}_{ij}$. Therefore, if the new variables are $N$, $N_i$ and $g_{ij}$, then
$$\pi^{ij}_{\mathrm{ADM}}=\pi^{ij}_{\mathrm{metric}} + \pi^{00}_{\mathrm{metric}} N^i N^j=\pi^{ij}_{\mathrm{metric}} + \pi^{00}_{\mathrm{metric}} \frac{g^{0i}g^{0j}}{(g^{00})^2},$$
and therefore
$$\{N,\pi^{ij}_{\mathrm{ADM}}\}=\left\{\frac{1}{\sqrt{g^{00}}}\ ,\ \pi^{ij}_{\mathrm{metric}} + \pi^{00}_{\mathrm{metric}} \frac{g^{0i}g^{0j}}{(g^{00})^2}\right\}=0.$$
Of course, $\pi^{00}=0$ in Dirac approach, so that it would mean that $\pi^{ij}_{\mathrm{ADM}}=\pi^{ij}_{\mathrm{metric}} $ on shell, or in terms of Lagrangian velocities, so that a direct and naive implementation of the change of variables in the degenerate Lagrangian would not be canonical in this sense. However, barring the possibility of different neglected boundary terms, the transition from metric to ADM variables is a canonical transformation if we employ the transformations like $\pi^{ij}_{\mathrm{ADM}}=\pi^{ij}_{\mathrm{metric}} + \pi^{00}_{\mathrm{metric}} N^i N^j$, not changing anything except indeed reparametrising the primary Lagrange multipliers in terms of canonical variables. This is yet another manifestation of the canonical Hamiltonian (\ref{CanHam}) being not unique for a degenerate system.

\section{Conclusions}

My main conclusion is that we should start being more serious about the constrained Hamiltonian analysis and stop taking it as just some sacred prescription which has to be taken with no deep thought. And even the simplest examples can demonstrate very interesting behaviour. Moreover, a very important point is that the adjectives of "physical" and "dynamical" are not the same. Both Maxwell electrodynamics and general relativity have only two dynamical modes each, but it is incorrect to say the same about their physical modes.

When it comes to our usual theories, they are well understood and the Hamiltonian formalism indeed works well even in the most naive ways. However, nowadays we are also studying many new and exotic fields theories, up to often classifying all possible models in a certain big class. In these situations, and with much less good physical intuition available yet, we must exercise even more care than ever before.

\end{document}